\documentclass[12pt,preprint]{aastex}

  \shorttitle{SGR\,1806-20 Giant Flare}
  \shortauthors{Boggs et al.}

\begin{document}

  \title{The Giant Flare of December 27, 2004 from SGR\,1806-20}

  \author{Steven E. Boggs\altaffilmark{1}, A. Zoglauer, E. Bellm, K. Hurley, R. P. Lin}

  \affil{
  Space Sciences Laboratory, University of California, Berkeley, CA 94720-7450, USA
  }

  \author{D. M. Smith}

  \affil{
  Department of Physics, University of California, Santa Cruz, CA 95064, USA
  }

 \author{C. Wigger, W. Hajdas}

  \affil{
  Paul Scherrer Institute, 5232 Villigen PSI, Switzerland
  }

  \email{boggs@berkeley.edu}

  \altaffiltext{1}{Department of Physics, University of California, Berkeley.}


  \begin{abstract}
The giant flare of December 27, 2004 from SGR\,1806-20 represents one of the most extraordinary events captured
in over three decades of monitoring the $\gamma$-ray sky. 
One measure of the intensity of the main peak is its effect on X- and $\gamma$-ray instruments. RHESSI, an instrument designed to study the brightest solar flares, was completely saturated for $\sim$0.5 s following the start of the main peak. A fortuitous alignment of SGR\,1806-20 near the Sun at the time of the giant flare, however, allowed RHESSI a unique view of the giant flare event, including the precursor, the main peak decay, and the pulsed tail. Since RHESSI was saturated during the main peak, we augment these observations with Wind and RHESSI particle detector data in order to reconstruct the main peak as well. Here we present detailed spectral analysis and evolution of the giant flare. We report the novel detection of a relatively soft fast peak just milliseconds before the main peak, whose timescale and sizescale indicate a magnetospheric origin. We present the novel detection of emission extending up to 17\,MeV immediately following the main peak, perhaps revealing a highly-extended corona driven by the hyper-Eddington luminosities. The spectral evolution and pulse evolution during the tail are presented, demonstrating significant magnetospheric twist and evolution during this phase. Blackbody radii are derived for every stage of the flare, which show remarkable agreement despite the range of luminosities and temperatures covered. Finally, we place significant upper limits on afterglow emission in the hundreds of seconds following the giant flare.

  \end{abstract}

  \keywords{gamma rays: SGR --- magnetars --- RHESSI}



  \section{Introduction}

The soft gamma repeater SGR\,1806-20 was discovered in 1979 \citep{lar86}, and 
has been studied intensively over the intervening two decades at X-ray, $\gamma$-ray, 
infrared, and radio wavelengths.  It has emitted over 450 soft $\gamma$-ray bursts, mostly
of short duration, during sporadic
active periods, and has been found to be a quiescent, variable X-ray source as well, emitting
up to $\sim$150 keV \citep{mer05c,mol05}.  Indeed, X-ray
observations of its periodic, quiescent component have provided some of the
best evidence for a magnetar-strength magnetic field \citep{kou98}, as first proposed
by \cite{dun92} and \cite{pac92}.  The infrared counterpart
to SGR\,1806-20 is a faint, highly obscured source, in keeping with its location
towards the Galactic center \citep{kos05,isr05}.  Presumably,
it is a lone neutron star, whose infrared intensity varies roughly in concert
with bursting activity and its quiescent X-ray flux.  There have been numerous
attempts to determine the distance to SGR\,1806-20 by various methods, leading
to estimates from 6.4 -- 9.8 kpc \citep{cam05} to 15.1 kpc \citep{mcc05,eik04,cor04,cor97}. 
In this paper, we will quote all energies and luminosities in terms of $\rm d_{10}=
(d/10 kpc)$.  Like SGR\,0525-66 and SGR\,1900+14, SGR\,1806-20 has emitted
a long duration, hard spectrum giant flare, whose flux at Earth greatly exceeded that of any other known
cosmic X-ray source \citep{hur05,maz05,mer05a,pal05}.
SGRs are not detectable quiescent radio emitters \citep{lor00}, 
but giant flares create transient radio nebulae which are observable for weeks \citep{fra99,gae05}.

In the magnetar model, magnetic dissipation, rather than rotation, provides the main energy source \citep{tho95,tho96}. 
Steady dissipation heats the neutron star surface and
powers the quiescent X-ray emission.  Localized crustal cracking causes short-duration, soft spectrum, and
relatively weak bursts during active periods.  Major crustal reconfigurations are thought to
be responsible for the rarer long-duration, hard spectrum, intense giant flares.

In 2004, SGR\,1806-20 underwent a period of intense activity.  The rate of small bursts peaked around
mid-year, in conjunction with the quiescent X-ray flux \citep{woo06}.  The spindown
rate, as evidenced by the frequency derivative, decreased.  This activity culminated in
the giant flare of December 27, 2004.  Unlike the case of the giant flare from SGR\,1900+14,
however, there was no sudden change to the spin frequency.  X-ray observations carried out
several months later, however, revealed a slower spin-down rate, a smaller pulsed fraction
for the quiescent emission, a different pulse profile, a softer spectrum, and a
decreased flux \citep{tie05,rea05}.  In the magnetar model, these changes
are attributed to a major reconfiguration of the neutron star's magnetic field.

In this paper, we present a detailed analysis of the \it Ramaty High Energy Solar Spectroscopic
Imager \rm (RHESSI) data on the giant flare, concentrating on the time-resolved energy spectra
of its various phases from 3\,keV to 17\,MeV.  By virtue of the high time and energy resolution and broad
spectral coverage of the measurements, we believe that these data present the most complete
spectral picture of this, or any other giant flare.  

\section{RHESSI Spectrometer}

RHESSI is an array of nine coaxial germanium detectors,
designed to perform detailed spectroscopic
imaging of X-ray and $\gamma$-ray emission (3\,keV - 17\,MeV) from
solar flares \citep{lin02}. Spectral resolution ranges from $\sim$1\,keV FWHM in the hard X-ray range, up to several keV in the MeV range. RHESSI imaging is
performed by two arrays of opaque 1-D grids, separated by 1.55 m, and co-aligned
with the nine detectors \citep{zeh03}. As the RHESSI spacecraft rotates (4.07\,s
period, axis aligned with the Sun) these grids modulate the count rate in
the detectors, allowing imaging through rotational modulation collimator techniques
\citep{hur02}. Thus, RHESSI has high
angular resolution (2.3$\,^{\prime\prime}$) in the 1$^\circ$ field of view of its optics.
However, the detectors themselves are unshielded, and are able to view transient sources from
the whole sky.
In addition, RHESSI sends down the energy and
timing information for each photon, allowing detailed timing measurements.

RHESSI observed both the precursor to and the giant flare from SGR\,1806-20
in their entirety, starting at 21:28:03.44 UT and 21:30:26.64 UT 
2004-12-27 at the spacecraft respectively. At the time of this event, SGR\,1806-20 was located 5$^{\circ}$ from the Sun, just outside the primary imaging field of view of the RHESSI instrument. At this angle from the solar direction, the shadow pattern of one front grid falls on the aligned bottom grid of a neighboring detector once per rotation, a fortuitous alignment that allows us to get a $\frac{1}{4}$-second ``snapshot'' of the direct spectrum, down to 3 keV, twice per RHESSI rotation period (4.07\,s). During the main peak of the flare the RHESSI spectroscopy detectors were saturated for $\sim$0.5\,s after the initial rise of the main peak, but observed the decay of the main peak and the 400-s long 
pulsed tail.

Fig.\,1 shows the 20-100\,keV lightcurve of the SGR\,1806-20 giant flare, from just before the precursor to after the end of the pulsed tail. In this plot and throughout this paper, we are quoting times t$_{26}$ in seconds relative to 21:30:26 UT 2004-12-27. (So the main peak starts at t$_{26}$ = 0.65\,s.) Since SGR\,1806-20 was located just outside of the main RHESSI field-of-view, this lightcurve is dominated by photons which have scattered into the detectors from other parts of the spacecraft or have been Compton reflected from Earth's atmosphere. Therefore, we do not attempt to use these events for spectral analysis below $\sim$0.2\,MeV. All of our hard X-ray spectra are derived using the snapshot data (with the exception of the fast peak), where RHESSI has a direct view of the flare. At higher energies, the RHESSI grids are more transparent and Earth reflection is negligible, allowing all photon events to be used for spectral analysis.

For the snapshot spectra we have used the on-axis RHESSI response matrices \citep{smi02}, which produce acceptable spectral fits to the snapshot data. For the RHESSI snapshot spectra analyzed in this paper we assume an absorption of $N_H = 6.69 \times 10^{22}$ cm$^{-2}$, which was measured for this source prior and subsequent to the giant flare \citep{mur94,mer05b,rea05}.

\section{RHESSI \& Wind Charged Particle Detectors}

During the intense main peak all X- and $\gamma$-ray instruments experienced some degree of saturation, making reliable reconstruction of the time history and energy spectrum difficult or impossible.  Many particle detectors, however,
are small, thin silicon detectors with very low effective areas for X- and $\gamma$-ray interactions.
Most of these detectors are usually impervious to $\gamma$-ray photons; however, due to the brightness of the main peak a number of particle instruments registered strong signals without saturating, allowing detailed reconstruction of the main peak \citep{hur05,ter05,sch05}. 

The RHESSI particle detector (PD), which is used for detecting SAA passages, is a small silicon detector of area 0.25 cm$^2$, 960 microns thick \citep{smi02}.
The RHESSI PD is instrumented to measure simple counting rates for all interactions above two threshold levels, $\sim$50\,keV and $\sim$620\,keV, with 0.125-s time resolution. The PD, which normally does not register even the brightest solar flares, measured a strong burst of counts from the main peak (Fig.\,2), but still well below saturation levels. Therefore, the PD gives us a very good measure of the incident count rates above these two thresholds. However, the limited PD data do not allow us to strongly constrain the shape of the $\gamma$-ray spectrum.

Therefore, we also analyzed the observations from the Wind 3D Plasma \& Energetic Particle Experiment \citep{lin95}. Wind  has  six double-ended  Solid State Telescopes (SSTs), five with two back-to-back 1.5\,cm$^2$, 300 micron thick silicon detectors  (called O and F, with 9 and 7 PHA channels, respectively), and one SST with a third, 1.5\,cm$^2$, 500 micron thick detector (T) in between. The  multi-channel analyzers covered the 20\,keV -- 11\,MeV range with various time resolutions between 12 and 96 s. All of the Wind detectors see a strong signal from the main peak, and they were all used in our analysis. The Wind SST data is saturated, but in a fairly benign way. The detectors and shaping electronics have a fast response (0.5\,$\mu$s shapers), and are well below saturation level. However, the analog-to-digital converters (ADC) for the telescopes are shared, and exceeded their maximum throughput during the main peak. When the ADC is busy processing an event, incoming events are thrown away, but without pileup (D. Curtis, private communication). Therefore, the Wind SSTs accurately sampled the spectral shape of the main peak, but not the overall normalization.

Between the Wind SSTs and the RHESSI PD we are able to reconstruct the input spectrum and the overall normalization, respectively. While our work focuses on the spectral analysis, other particle detector observations have been able to reconstruct the main peak lightcurve with much higher time resolution \citep{ter05, sch05}. The summary of these spectral results was presented in \cite{hur05}. Here we present more details of the analysis method.

\section{Stages of the Giant Flare}

For the purpose of our analysis, the giant flare from SGR\,1806-20 can be divided into six separate stages, illustrated in Fig.\,1. There is a precursor flare (i) 142 seconds prior to the main peak, followed by a quiescent preflare period (ii). Immediately prior to the main peak, a fast peak (iii) occurs which lasts merely 2.5\,ms. The main peak itself (iv) lasts $\sim$0.5\,s. Subsequent to the main peak, there is a brief $\sim$60\,s decay period dominated by strong nonthermal emission (v), followed by the characteristic pulsed tail (vi) lasting 400 seconds. Finally, there is the postflare period with the potential for afterglow emission (vii). While RHESSI was saturated $\sim$0.5\,s during the main peak, it had an excellent view of SGR\,1806-20 during the rest of these stages. Here we present a detailed analysis of each stage of this spectacular event.

\subsection{Precursor}

RHESSI observed a precursor burst during 21:28:03.44-21:28:04.49 UT,
142\,s before the main peak of the giant flare, with a 
peak count rate in the spectroscopy detectors of $\sim$30,000 cnt\,s$^{-1}$.
The relatively long duration (1\,s) and nearly flat lightcurve of this burst (Fig.\,3) distinguish it from more common SGR bursts with typical durations of $\sim$0.1\,s.  We 
see a rise time for the precursor of 27 ms, and a fall time of 110 ms. 
RHESSI caught this precursor during one of its serendipitous spectral snapshots (Fig.\,3), allowing us to get a good spectrum of it. The 3-250\,keV spectrum (Fig.\,4) is well fit by a single blackbody component, assuming $N_H = 6.69 \times 10^{22}$ cm$^{-2}$, with $kT = 10.4 \pm 0.3$ keV ($\chi_{\nu}^2 = 1.06$, 75 dof). By comparison, both a simple power law model ($\chi_{\nu}^2 = 4.02$, 75 dof) and a thermal bremsstrahlung model ($\chi_{\nu}^2 = 2.03$, 75 dof) give unacceptable fits to the precursor. The spectrum shows no evidence for nonthermal emission in addition to the simple blackbody. Using this spectral snapshot, we can estimate the time-integrated blackbody fluence of this precursor to be $(3.2 \pm 0.5) \times 10^{-5}$ erg\,cm$^{-2}$, implying an energy of $3.8 \times 10^{41}$ $d_{10}^2$ erg. Note, this spectral fit and fluence differ significantly from the results reported in \cite{hur05} due to an error in our preliminary analysis of this precursor. (This error did not affect the rest of our preliminary analysis.) This current analysis resolves most of the apparent discrepancy between our previous precursor fluence and that reported from ACS/INTEGRAL \citep{mer05a}.

\subsection{Preflare}

Following the precursor, SGR\,1806-20 appears relatively quiescent for 142\,s until the main peak of the giant flare. While there is no obvious evidence in the lightcurve for emission during this period (Fig.\,1), RHESSI snapshot spectra allow us to search for emission with better sensitivity than the lightcurves alone. We see no evidence of emission from SGR\,1806-20 during this quiescent period. The 3$\sigma$ upper limit on the 3-250\,keV flux is $1.4 \times 10^{-7}$ erg\,cm$^{-2}$\,s$^{-1}$, corresponding to a source flux below $1.7 \times 10^{39}$ $d_{10}^2$ erg\,s$^{-1}$.

\subsection{Fast Peak}

There was somewhat less than one millisecond between the time that
the first signs of the main peak became detectable above background by RHESSI and the time the instrument went into saturation.  Fig.\,5 shows
the raw count rate at the start of the rise.  We fit the early
data where the instrument livetime was $>$ 90\% with an exponential
growth curve, shown as well in Fig.\,5.  Normally, rear segment response at low energies on RHESSI is dominated by Compton reflection from Earth's atmosphere, but during the fast rise these photons would not have had a chance to reach the instrument, and therefore all segments are included in the lightcurve of Fig.\,5. The best-fit e-folding 
time constant is $0.38 \pm 0.04$\,ms. This rise time is comparable to the $\sim$ 0.3\,ms rise time reported for Swift-BAT before saturation \citep{pal05}, but an order of magnitude faster than the 4.9\,ms rise time reported by particle detector observations that did not saturate \citep{sch05}. 
However, we can also see in Fig.\,5 that RHESSI partly recovers during the $t_{26} = 638 - 639$\,ms period, suggesting that the onset of the main peak was not smooth, but had a significant drop in the incident rate $\sim$ 2.5\,ms after the start of the rise. This drop is also evident in the Swift-BAT lightcurve at this same time \citep{pal05}.
One explanation of the apparent discrepancy between the rise time measurements is that RHESSI and Swift-BAT are characterizing the rise time of this initial fast peak, while the particle detectors are characterizing the main peak itself, following this initial fast peak.

The fast rise time and short duration of this fast peak suggest that it represents a separate physical mechanism from the main peak itself. This idea is also supported by the spectrum of the fast peak.
Fig.\,6 shows the RHESSI count spectrum during the $t_{26} = 636.2-636.7$\,ms period. RHESSI did not get a direct snapshot spectrum during this period, so only a rough spectral analysis is possible. We can compare this count spectrum with the measured precursor spectrum with $kT = 10.4$\,keV (solid line, Fig.\,6). Here we plot the precursor count spectrum after the snapshot period (Fig.\,4) for a direct comparison between the two count spectra outside snapshot periods. This fast peak spectrum appears harder than the precursor. However, this spectrum is much softer than the $kT = 175$\,keV blackbody (dashed line, Fig.\,6) that we measure for the main peak itself (next section). A $kT \sim 20$\,keV blackbody convolved through the rotation-averaged instrumental response gives a reasonable match to the measured count spectrum (dot-dashed line, Fig.\,6). If we assume this component is a 20\,keV blackbody, we can derive a rough fluence for this 0.5\,ms period during the fast rise of $6.6 \times 10^{-7}$ erg\,cm$^{-2}$.

\subsection{Main Peak}

The Wind SST F \& O count spectra of the main peak are shown in Fig.\,7. As discussed in \S\,3, the Wind SST detectors measured the main peak spectral shape, but not the overall normalization. 
We concentrated our efforts on the F \& O spectra during this analysis --- the combined spectra of six detectors each. We exclude the so-called FT and OT coincident spectra due to the poorer statistics in this data mode, and uncertainties in the coincidence trigger criteria. 
In addition, the lowest bin in the F spectrum had to be excluded due to uncertainties concerning trigger threshold effects.  
In order to determine the best-fit spectral model, we developed a simulation mass model of the Wind 3D experiment, including a detailed detector and housing model, and a rough spacecraft model. 
Special care has been taken to correctly model the passive material on all direct paths from the magnetar to the active detector materials and the immediate surrounds of the detectors to correctly account for absorptions and scatters.
The mass model itself was developed in MEGAlib \citep{zog06} to allow both GEANT3 and GEANT4 Monte Carlo simulations in order to cross check results. 
Of specific concern was also the correct handling of electron tracks from Compton and pair interactions given the thin F \& O detectors. 
However, since no calibration exists for this detection mode, some uncertainties remain.
Given that most photons above roughly 50 keV interacting in one of the SST detectors will Compton scatter out of the instrument, the photon response matrix is strongly non-diagonal. 
As consequence, we did not attempt to create and invert a photon response matrix and produce model-independent spectra, but rather compared measured and simulated count spectra directly to determine the best match to the overall measured spectral shapes.

We attempted to reproduce the observed count rate distributions with power law, thermal bremsstrahlung, and blackbody spectral models. We varied the input spectral parameter for each model (photon index, temperature) over a range of values, adjusting the overall normalization to best match the observed count spectrum. We verified that the range of spectral values bracketed the best-fit value for each spectral model. The best-fit power law and bremsstrahlung spectra were strongly rejected ($\chi_{\nu}^2 = 4.2$ and $6.9$, 10 dof), and only the blackbody with $kT = 175 \pm 25$ keV provided an acceptable fit ($\chi_{\nu}^2 = 1.0$, 10 dof).

Given the best-fit spectral model from Wind SSTs, we can use the RHESSI PDs (Fig.\,2) to constrain the overall fluence. The RHESSI PD was also modeled in GEANT3, assuming the best-fit blackbody model derived with the Wind SSTs above. The RHESSI PD data yield an overall fluence of $(1.36 \pm 0.35)$ erg\,cm$^{-2}$ (for the blackbody integrated over all energies), implying an isotropic energy release in the main peak of $1.6 \times 10^{46}$ $d_{10}^2$ erg. Given the RHESSI PD time resolution, the peak flux in the first 0.125\,s was $1 \times 10^{47}$ $d_{10}^2$ erg\,s$^{-1}$.

While the main peak is evident in only two RHESSI PD 0.125-s time bins, it shows clear indication of softening between these two intervals (Fig.\,2). Assuming an input blackbody spectrum, we can use the two PD count rates to characterize the blackbody temperature independent of the Wind results. In the first and second time bins, the RHESSI PD characterizes the temperatures as $kT = 230 \pm 20$ keV and $kT = 170 \pm 5$ keV respectively.

\cite{haj05} performed a similar analysis on the IREM radiation environment monitor on the INTEGRAL satellite. IREM has three Si-diodes with crude spectral response, consisting of count rates read out every 60\,s for events above six energy thresholds spanning 85\,keV to above 3\,MeV. The IREM spectral analysis results in a cooling blackbody with $kT = 230 \pm 50$ keV and an overall fluence of $(0.97 \pm 0.50)$ erg\,cm$^{-2}$, consistent with our analysis above. By contrast, \cite{maz05} performed spectral analysis of the $\gamma$-ray signal measured by the Russian spacecraft Coronas-F. This spacecraft was occulted by the Earth at the time of the main peak, but observed the event reflected off of the Moon. They simulated the response of scattering off the Moon, then folded this response with their detector response matrix. Their best-fit model for this spectral analysis is a powerlaw function ($\Gamma = 0.7$) with exponential cutoff ($E_o = 800$\,keV). However, from our own spectral analysis of the Wind SST data, we can rule out this spectral model with very high confidence ($\chi_{\nu}^2 > 10$, 11 dof). Therefore, given the quality of the Wind SST data, the consistency with the limited RHESSI PD data, and the consistent results with the IREM data, we are confident of the $kT \sim 200$\,keV cooling blackbody spectrum of the main peak.

\subsection{Peak Decay}

When they came out of saturation within 1\,s after the start of the main peak, the RHESSI
spectroscopy detectors were measuring a peak count rate of
$\sim$280,000 cnt\,s$^{-1}$. During the first few seconds, RHESSI recorded a dynamic and complex spectrum. What stands out most in the lightcurve (Fig.\,8) is the pulsed tail, composed of both thermal blackbody emission and a nonthermal power law emission. Both of these spectral components are present immediately when RHESSI comes out of saturation, with the earliest snapshot at t$_{26}$ = 2.0\,s. These two components, and their evolution throughout the pulsed tail, are discussed below.

An additional component present as RHESSI emerged from saturation consists of strong emission extending up to 17 MeV, the upper limit of RHESSI's energy band. To our knowledge, this is the highest energy to which this or any other SGR flare has been observed. Fig.\,9 shows the 0.4-10 MeV lightcurve for the first 300 seconds after the giant peak, including our fit to the background rate before and after the giant peak. RHESSI observes excess emission in the MeV range for $\sim$60\,s after the giant peak, which is better modeled as a power law decay than an exponential decay. Fitting the background-subtracted lightcurve with a function of the form $\propto t^{-a}$ yields a best-fit index $a = 0.68 \pm 0.04$ ($\chi_{\nu}^2 = 1.63$, 39 dof). An MeV component was previously reported in the pulse-averaged spectra from Konus-Wind observations to $\sim$10\,MeV \citep{maz05}, though we do not see any significant MeV emission after t$_{26}$ $\sim$ 100\,s. 

We verified that the high-energy component is not created by low-energy
photons from the SGR arriving simultaneously with high-energy background
photons and creating an artificial ``pileup'' component at high energies.
During the period of high-energy emission in the peak decay, the
detector livetime averaged around 96\% in the RHESSI rear segments. 
Using pileup-modeling software based on RHESSI ground calibrations and
solar flare observations, we find that the high-energy contribution of
pileup should be about a factor of 10 lower than the
high energy component observed.

For the t$_{26}$ $=$ 1.71-22.06\,s period, the 0.4-15\,MeV spectrum (Fig.\,10) can be fit by a power law ($\frac{dN}{dE} \propto E^{-\Gamma}$) of photon index $\Gamma = 1.43 \pm 0.06$, with an integrated fluence of $(9.8 \pm 0.1)\times 10^{-5}$ erg\,cm$^{-2}$. For this fit, we excluded the band around the 0.511\,MeV background line, which is difficult to model and subtract properly for this transient event. The power law model for this MeV component, shown in Fig.\,10, is a good fit above 0.4\,MeV ($\chi_{\nu}^2 = 0.86$, 30 dof). There is no sign of a turnover in this spectrum up to 17\,MeV. Adding an exponential cutoff to the models marginally worsens the spectral fit, and results in a cutoff energy $>$50\,MeV, well above the RHESSI energy range.

Below 0.4\,MeV, Fig.\,10 shows a strong excess above our simple power law model. 
This excess is partly due to photons from SGR\,1806-20 that scattered in the earth's atmosphere before reaching the spectroscopy detectors, and due in smaller part to a softer power law index at lower energies as revealed in the snapshot spectra (Sec.\,4.6). While the snapshot spectra themselves show no sign of a change in the spectral index ($\sim$2.5) below 250\,keV, the higher energy data in Fig.\,10 show that the spectral index steepens ($\sim$1.5) at higher energies. The exact energy of this spectral break is only weakly constrained by the RHESSI data to be $0.5 \pm 0.2$\,MeV.

\subsection{Pulsed Tail}

After the main peak of the giant flare, RHESSI recorded a series of 51 
pulsations with a period of 7.56\,s (Fig.\,1), similar to the INTEGRAL, KONUS, and 
Swift-BAT observations \citep{mer05a,maz05,pal05}. The pulse profile shows evidence for 
both spectral variations throughout the pulse, and evolution of the 
pulse shapes throughout the decay. The 20-100\,keV pulse profiles 
show 3-4 peaks in their structure.

In Table\,1 we present the phase-integrated spectral evolution of the pulsed tail as seen through the snapshot spectra. Data were combined over the time periods presented, and fit in the 3-250\,keV band (assuming $N_H = 6.69 \times 10^{22}$ cm$^{-2}$) to determine the best spectral model. For these pulse-integrated spectra, the best-fit model is a two-component blackbody plus power law (photon index $\Gamma$). Thermal bremsstrahlung and power law models give unacceptable fits, both alone and combined as a two-component model. A blackbody model alone gives marginal fits, improved significantly with the addition of the nonthermal powerlaw component. We do not see any significant sign of an exponential cutoff in this power law component below 250\,keV, the upper end of our snapshot data. After t$_{26}$ $\sim$ 246\,s, the power law photon index is not strongly constrained in the individual 40-s spectra though the component is still significant; therefore, we have fixed the index at its best-fit value for this time period, $\Gamma = 2.1$. An example spectral fit is shown in Fig.\,11, the first snapshot spectrum after RHESSI emerged from saturation.

In our preliminary analysis we could not conclusively distinguish between the thermal bremsstrahlung and blackbody models, which both gave marginal fits \citep{hur05}. The addition of the absorption column and the power law component improved the blackbody fits to the point of strongly distinguishing the models. 

The blackbody component appears to be present in the tail emission immediately after RHESSI comes out of saturation, with an initial temperature $kT = 11.5$\,keV, which drops steadily during the evolution of the pulsed tail (Table\,1). The total integrated fluence of the blackbody component of the pulsed tail is $(2.6 \pm 0.2) \times 10^{-3}$ erg\,cm$^{-2}$.

The power law component is also present immediately after RHESSI comes out of saturation, and lasts throughout the pulsed tail phase. The photon index, initially $\Gamma = 1.71$, appears to soften to $\Gamma = 2.7$ then harden to $\Gamma = 2.1$ over the evolution of the tail, but it is not clear that this evolution is strongly significant. The total integrated fluence of the nonthermal component for the pulse tail phase, in the 3-100\,keV band, is $(2.9 \pm 0.5) \times 10^{-3}$ erg\,cm$^{-2}$.

Combining these two components, the total fluence measured from the pulsed tail is $(5.5 \pm 0.6) \times 10^{-3}$ erg\,cm$^{-2}$, implying an energy release of $6.7 \times 10^{43}$ $d_{10}^2$ erg, roughly equally divided between thermal and nonthermal emission.

Fig.\,12 shows the average 7.56-s pulse shape, 20-100\,keV, integrated over the pulsed tail. The pulse profile is dominated by 3-4 separate peaks, with an overall large pulse fraction. Fig.\,12 also shows the phase-resolved best-fit blackbody temperature, which varies throughout the pulse. During the phase-resolved profile, the ratio of 3-100\,keV power law flux to the total blackbody flux stays flat, with the energy emission nearly equally divided between the two components.

\subsection{Afterglow}
ACS/INTEGRAL observations of the SGR\,1806-20 giant flare showed excess counts in the period following the pulsing phase, 
t$_{26}$ $\sim$ 400-4000\,s, peaking around t$_{26}$ $\sim$ 600-800\,s. This excess was interpreted as afterglow hard X-ray emission from the SGR \citep{mer05a}. The ACS observations suggest that the integrated fluence in this afterglow emission is comparable to the integrated emission in the pulsing tail itself. RHESSI had a direct view of SGR\,1806-20 for 600\,s following the pulsing tail before the satellite moved behind Earth's shadow. We have searched the RHESSI snapshot spectra during this period for evidence of afterglow emission. For t$_{26}$ $\sim$ 400-1000\,s, we can set a 3$\sigma$ upper limit on the 3-200\,keV fluence from SGR\,1806-20 of $1.4 \times 10^{-4}$ erg\,cm$^{-2}$. This upper limit is a factor of 60 below the fluence expected based on the ACS observations. Given our snapshot spectra, we are effectively chopping between source and background. Therefore, our analysis is not sensitive to potential background variations as is the ACS lightcurve analysis. Activation of the BGO crystals comprising the INTEGRAL ACS detectors was considered as a possible explanation for the excess counts, but it is not clear that this would be consistent with the observed lightcurve of the ACS afterglow (Mereghetti, private communication).

\section{Discussion}

The precursor may hold some tantalizing clues to the origin of this giant flare. 
The unusual nature of this burst, and its occurence soon before the main peak, suggest a direct connection between the two events.
Indeed, a precursor was also present for the giant flare from SGR\,1900+14 \citep{hur99}.
The average luminosity during the precursor was $4 \times 10^{41}$ $d_{10}^2$ erg\,s$^{-1}$, corresponding to $\sim$2000\,L$_{Edd}$, which is typical for SGR bursts. In Fig.\,13 we show the phase of this precursor relative to the subsequent pulsed tail -- the precursor occurs during one of the lowest phases of the pulse profile. 
Our measured rise time of this precursor, 27\,ms, is much longer than the $\sim$0.03\,ms timescale expected for magnetospheric realignment, but consistent with the expected 10\,ms timescale for crustal slipping after cracking \citep{tho01}. This timescale is strong evidence that the precursor originated from a fracture propagating in the crust of the neutron star, just as in the main peak itself \citep{sch05}. The total energy of this precursor, $3.8 \times 10^{41}$ $d_{10}^2$ erg, is comparable to the maximum elastic potential energy the magnetar can store in its crust before cracking, $\sim$10$^{42}$ erg \citep{tho01}, which might suggest that the precursor corresponds to a global cracking and realignment of the crust. However, the relatively long 1-s duration of the precursor (5$\times$ the main peak of the giant flare itself) and the multi-peaked lightcurve suggests instead that the precursor is being powered by repeated injections of energy by realignment of the magnetic field in the core, with a typical timescale on the order of 200\,ms \citep{tho01}, similar to the main peak itself \citep{sch05}. Given the energetics, this scenario suggests that the precursor corresponds to an energetically small crustal fracture, followed by repeated energy injections from a relatively small realignment of the core. This scenario is supported by two more pieces of evidence. First, our precursor spectrum is purely blackbody. If there were significant twisting of the magnetosphere we would have expected a nonthermal tail \citep{tho02}. Second, the drop in flux by over a factor of 200 after the precursor (during the preflare period) also suggests that the magnetosphere was not significantly twisted by the precursor event itself.

We measured the peak flux in the first 0.125\,s of the main peak to be $1 \times 10^{47}$ $d_{10}^2$ erg\,s$^{-1}$, an astounding $10^9$\,L$_{Edd}$. In our previous paper we showed that this luminosity and the measured temperature of $kT \sim 200$ keV are consistent for blackbody emission from a spherical surface of radius $R \sim 10$ km (see below). We have measured the isotropic energy release in the main peak of $1.6 \times 10^{46}$ $d_{10}^2$ erg. This is an awesome amount of energy for this source. For comparison, given the 7.56\,s period the rotational kinetic energy of SGR\,1806-20 (the power supply for radio pulsars) is on the order of $E_{spin} = \frac{1}{2}I\Omega^2 = 3 \times 10^{44}$ erg. Given that the energy release in the main peak is two orders of magnitude greater than this rotational energy of the star, it is even more amazing that there was no measurable jump in the spin frequency after the giant flare \citep{woo06}.

The energy release of the main peak is comparable to the maximum energy that could be stored in a twisted magnetosphere, thus energetically this giant flare would be consistent with global magnetospheric untwisting \citep{hur05}. However, in this scenario one would expect a growing spin-down rate before the giant flare, whereas it was actually decreasing in the months prior, and a significant drop in the spin-down rate after the giant flare, likely larger than the observed decreased which appears consistent with the trend before the giant flare \citep{woo06}. Thus we are left with the conclusion that this giant flare represents a large-scale crustal instability in the star, driven by the unwinding of the toroidal field inside the core \citep{tho01}. The main peak taps only a fraction of the $10^{48}$ - $10^{49}$ erg of magnetic energy stored in the core of the SGR.

This conclusion appears consistent with the strong nonthermal emission during the peak decay and throughout the pulsed tail, which indicates large magnetospheric twisting during the main peak \citep{tho02}.
The nonthermal emission during the pulsed tail, with spectral indices $\sim$ 2.1-2.7 and extending $>$250\,keV, is likely easily explained by electron cyclotron scattering within an extended corona, which can reach photon energies $\geq$100\,keV \citep{tho02}.
However, the second nonthermal component during the peak decay, with a spectral index of 1.43 and extending $>$17\,MeV with no sign of a spectral cutoff, is more difficult to explain. It clearly must derive from a separate mechanism than the electron cyclotron scattering. Ion cyclotron scattering is only expected to extend into the tens of keV range \citep{tho02}. We speculate that this component perhaps arises from a highly extended corona, driven by the hyper-Eddington luminosities, where synchrotron emission is no longer efficient at cooling the electrons \citep{fer01}. This scenario seems consistent with the lack of a clear pulsation from the SGR for over a full rotation period, nearly 10\,s, following the main peak.

Fig.\,13 shows that there is significant evolution of the pulse profile over the course of the pulsed tail. Each panel shows the average pulse profile integrated over ten consecutive pulses. Especially significant is the increase of the peak at phase $\sim$0.65 through the evolution of the tail. Evolution of the pulse profile demonstrates continuing magnetospheric realignment during the course of the pulsed tail. \cite{fer01} have studied a similar phenomenon in the pulse profile of SGR\,1900+14 following a giant flare.

The phase-averaged luminosity during the 400\,s pulsed tail corresponds to 300-2000\,L$_{Edd}$. In our previous paper, we showed that the pulse-averaged 
20-100\,keV flux lightcurve of the pulsed tail is well modeled by the trapped fireball model of 
\cite{tho01}, where flux $\propto (1-\frac{t}{t_{evap}})^{(\frac{a}{1-a})}$, with an evaporation time 
$t_{evap} = 382 \pm 3$ s, and index $a=0.606 \pm 0.003$ \citep{hur05}. This index is physically significant, being close to the expected value $a = \frac{2}{3}$ for a homogeneous, spherical trapped fireball \citep{tho01}.
Within this trapped fireball model, we can derive a rough bound on the magnetic field by requiring the magnetic field to be strong enough to confine energy radiated by the trapped fireball,
$B_{dipole} > 1.7 \times 10^{14} (\frac{\Delta R}{10 km})^{\frac{-3}{2}} [(1+\frac{\Delta R}{R})/2]^3 d_{10}$\,G \citep{tho95}.

We observed blackbody components, of different temperatures and luminosities, from the precursor, main peak, peak decay, and pulsating tail. For a source distance d, and surface gravitational redshift z, we can convert measured luminosities and temperatures into effective blackbody radii of the emission regions, $R = (\frac{L}{\sigma T^4})^{\frac{1}{2}} (\frac{d}{1+z})$. In Fig.\,14 we present our derived blackbody radii for the various stages of the giant flare, assuming $d = 10$\,kpc, and $z = 0.30$ for a neutron star ($\frac{M}{R} \sim \frac{1.4 M_{Sol}}{10 km}$). 
The first remarkable feature about these radii is that they all roughly agree throughout the stages of this giant flare, from the precursor (14 km) and the main peak (18 km), through the main peak decay (17 km) and the average throughout the pulsing tail (11 km). These agree remarkably well given the variation in temperature and luminosity throughout. In addition, we can turn this around to point out that, in terms of the uncertain distance to SGR\,1806-20, $d < 10$\,kpc would be more consistent with a canonical neutron star radius of $R \sim 10$\,km than $d = 15$\,kpc if the thermal emission is originating from the stellar surface as opposed to an extended corona.

The fast peak just prior to the main peak remains a bit of a puzzle. The average luminosity during the 0.5\,ms before RHESSI saturates was $1.6 \times 10^{43}$ $d_{10}^2$ erg\,s$^{-1}$, corresponding to $\sim 10^5$\,L$_{Edd}$. The peak in the lightcurve and the relatively soft spectrum clearly distinguish this as separate from the main peak itself. However, the rise time (0.4\,ms) and duration (2.5\,ms) of this fast peak are too fast for global crustal slippage or realignment of the core field \citep{tho01}. These times could imply a very localized crack in the crust ($\sim$ 0.4\,km), but this fast peak is much shorter than typical SGR flares (and the precursor), which are attributed to the same physical mechanism. 
Furthermore, if we assume the spectrum really corresponds to a blackbody of $kT \sim 20$\,keV, then we can also derive a blackbody radius of roughly 21\,km for this stage of the flare, i.e. a global event, presumably inconsistent with a localized crustal crack. 
If we divide this radius by the rise time we can derive a lower limit on the thermal diffusion speed, corresponding to $\sim$0.2\,c. This fast peak appears to only be consistent in timescale and sizescale with a realignment of the magnetosphere immediately before the main peak itself. While the energy of this fast peak is small compared to the main peak, its fast timescale and its occurrence milliseconds prior to the main peak suggest it plays some critical role in the giant flare.

The giant flare of SGR\,1806-20 represents one of the most outstanding events in X-ray and $\gamma$-ray astronomy over the past three decades, even when compared to the giant flares of SGR\,0525-66 and SGR\,1900+14. Even though this giant flare presents an extreme energy output from the SGR, the basic energetics and timescales involved are well understood in term of the magnetar model. Indeed, given the spin energy of the star and the extraordinary super-Eddington luminosities of the giant flare, SGR\,1806-20 has once more presented very strong evidence in favor of the magnetar model.

  \acknowledgments
  The authors are grateful to G. Hurford, H. Hudson, and S. Krucker for useful discussions. KH is grateful for support under the NASA Long Term Space Astrophysics program grant NAG5-13080.

  \clearpage

  \begin{figure}
  \epsscale{1.00}
  \plotone{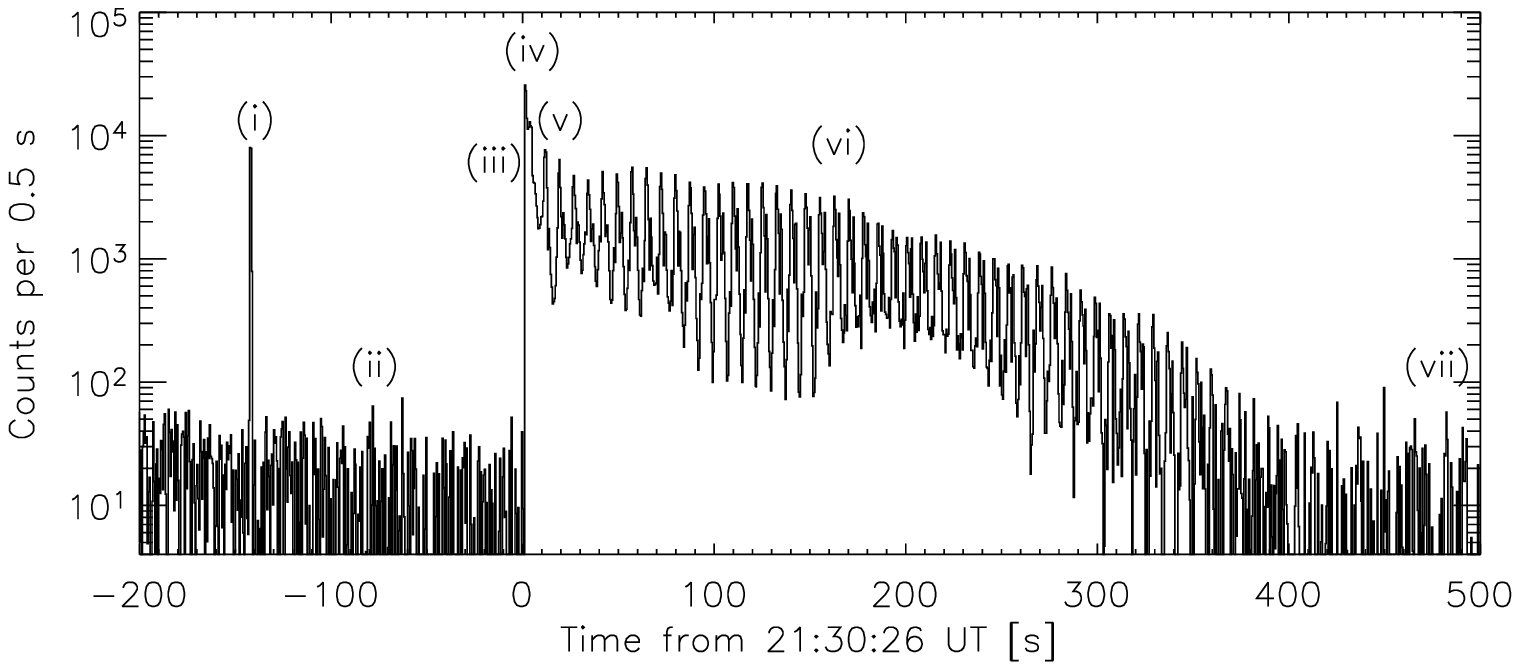}
  \caption{The RHESSI 20-100\,keV background-subtracted time history of the giant flare, plotted with 0.5-s resolution. The main peak begins at 0.64\,s, where the RHESSI detectors are saturated and effectively dead. RHESSI recovered $\sim$0.5\,s later to observe the rest of the giant flare in detail. In this paper we analyze six separate stages of this flare: (i) precursor, (ii) preflare, (iii) fast peak, (iv) main peak, (v) peak decay, (vi) pulsed tail, and (vii) afterglow. In this energy range, the time history is modulated by RHESSI's 4-s spin period; otherwise, there are no other long-term induced modulations (instrument repointing, saturation, etc.). Therefore, other than the main peak this represents the true lightcurve of the giant flare event. \label{fig1}}
  \end{figure}

  \clearpage

  \begin{figure}
  \epsscale{0.70}
  \plotone{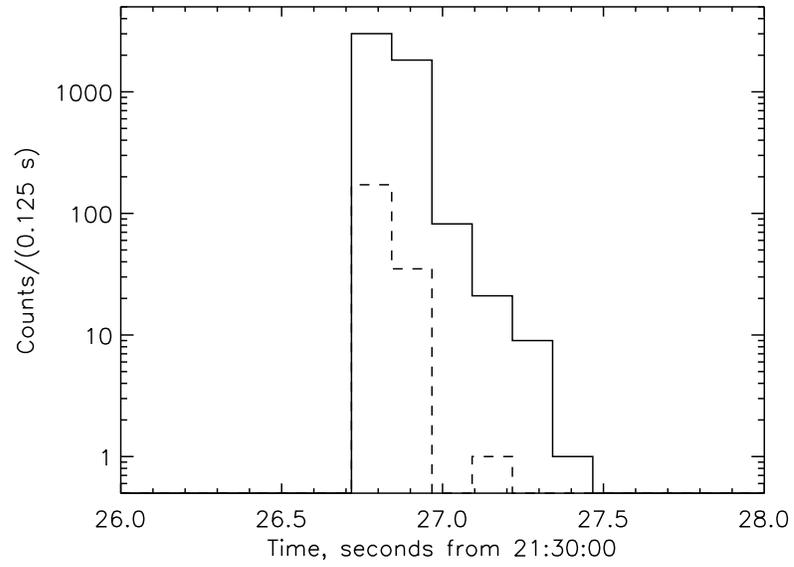}
  \caption{The RHESSI particle detector
was able to measure the incident flux with 0.125-s time resolution
in two energy channels determined by thresholds in the electronics:
$>$50 keV, and $>$620 keV. These data indicate significant emission above
620 keV for $\sim$0.25 s, during the main peak, and softening of the spectrum
during its evolution. \label{fig2}}
  \end{figure}

  \clearpage
  \begin{figure}
  \epsscale{0.80}
  \plotone{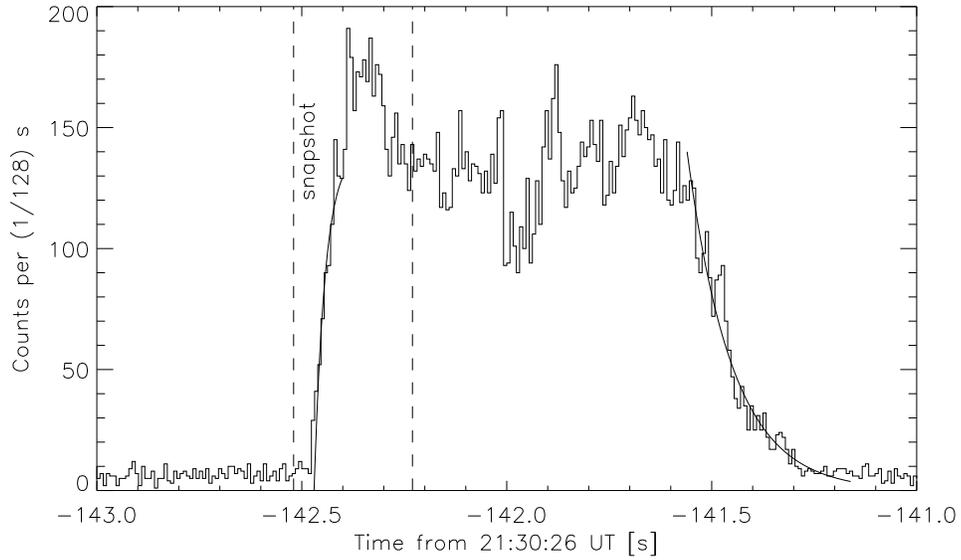}
  \caption{Precursor 20-100\,keV lightcurve, in $\frac{1}{128}$\,s time bins. The precursor has an exponential rise time of 27\,ms, and a fall time of 110\,ms (solid lines). The period of the precursor snapshot spectrum (Fig.\,4) is shown by the dashed lines. \label{fig3}}
  \end{figure}

  \clearpage
  \begin{figure}
  \epsscale{0.80}
  \plotone{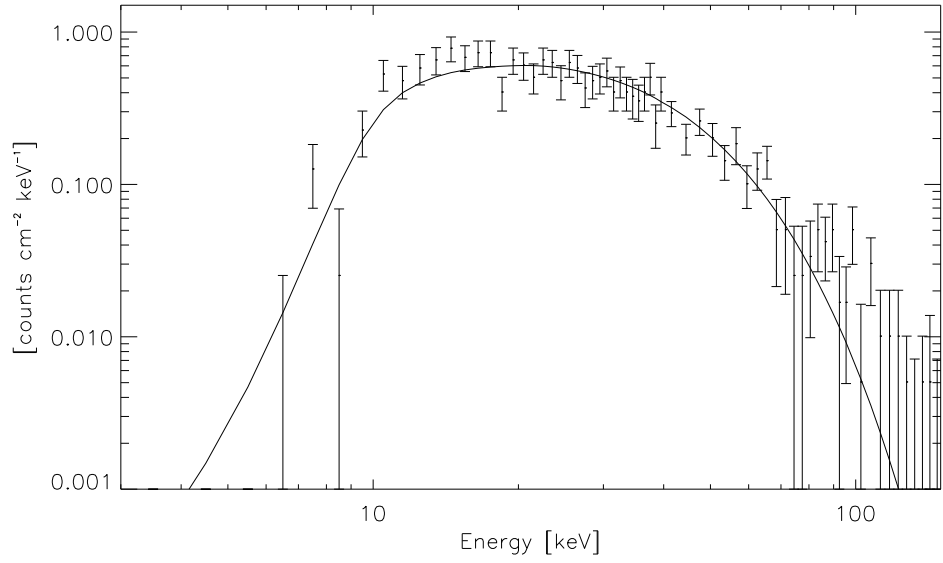}
  \caption{RHESSI snapshot spectrum of the precursor, 3-150\,keV. This spectrum is background-subtracted, and the solid line shows the best-fit absorbed blackbody spectrum convolved through the RHESSI response matrix. \label{fig4}}
  \end{figure}

\clearpage
  \begin{figure}
  \epsscale{0.80}
  \plotone{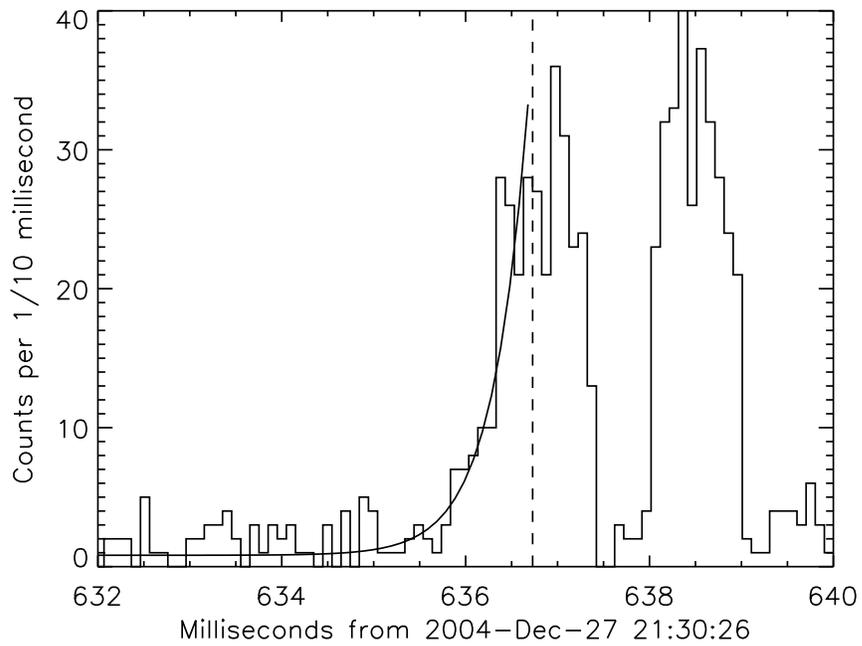}
  \caption{Total RHESSI count rate during the rising edge of the
  giant flare. Beyond the dashed line, instrument deadtime
  becomes significant; in fact, the peak in count rate
  at 638.5 ms represents a dip, not a peak, in
  the true flux from a partial recovery in livetime.  The instrument is almost completely
  paralyzed during the periods of low count rate after 637 ms.\label{fig5}}
  \end{figure}

\clearpage
  \begin{figure}
  \epsscale{1.00}
  \plotone{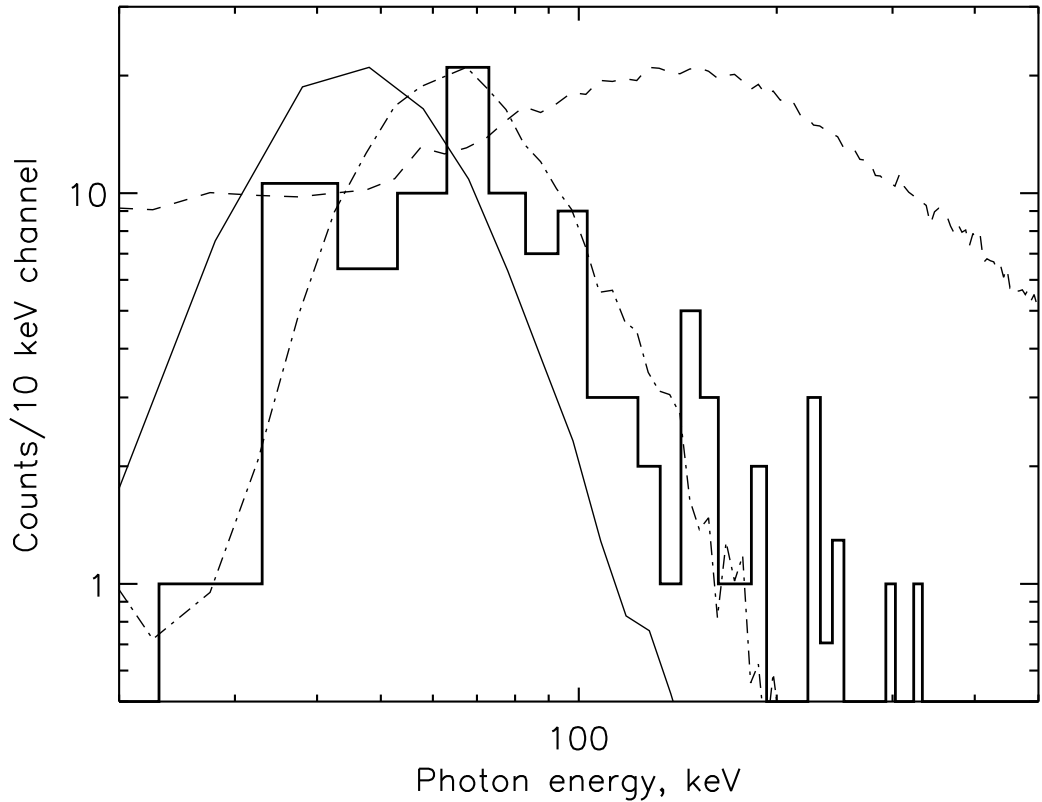}
  \caption{RHESSI count spectrum during the fast peak just prior to the
  main peak. RHESSI did not get a direct snapshot spectrum during this
  period, so only a rough spectral analysis is possible. The measured precursor spectrum (outside of the snapshot period) is shown for comparison (solid line). Also shown for comparison are two blackbody spectra convolved with the instrument response matrix: $kT = 175$\,keV corresponding to the main peak temperature (dashed line), and $kT = 20$\,keV (dot-dashed line). The normalizations on the three comparison spectra are arbitrary.
 \label{fig6}}
  \end{figure}

  \clearpage
  \begin{figure}
  \includegraphics[angle=270,scale=0.60]{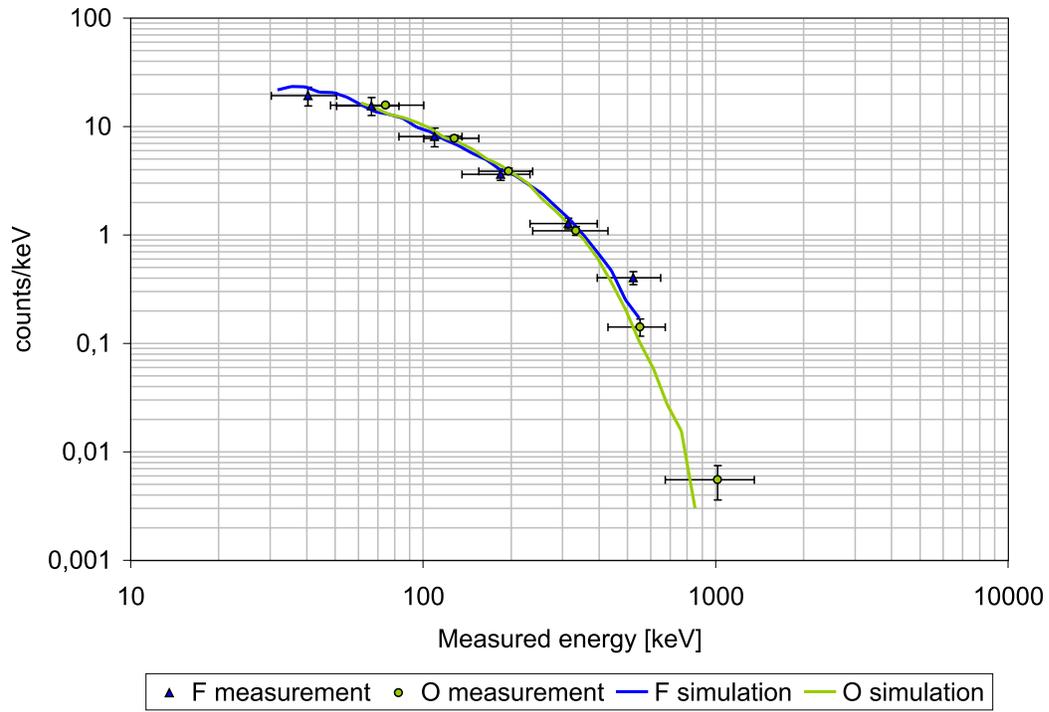}
  \caption{Count spectra of the main peak of the Wind SST F and O detectors. The crosses show the measured spectra and the lines the best-fitting 
simulated spectra, a blackbody with the temperature 175$\pm$25 keV ($\chi_{\nu}^2  
= 1.0$, 10 dof). \label{fig7}}
  \end{figure}

  \clearpage
  \begin{figure}
  \epsscale{0.90}
  \plotone{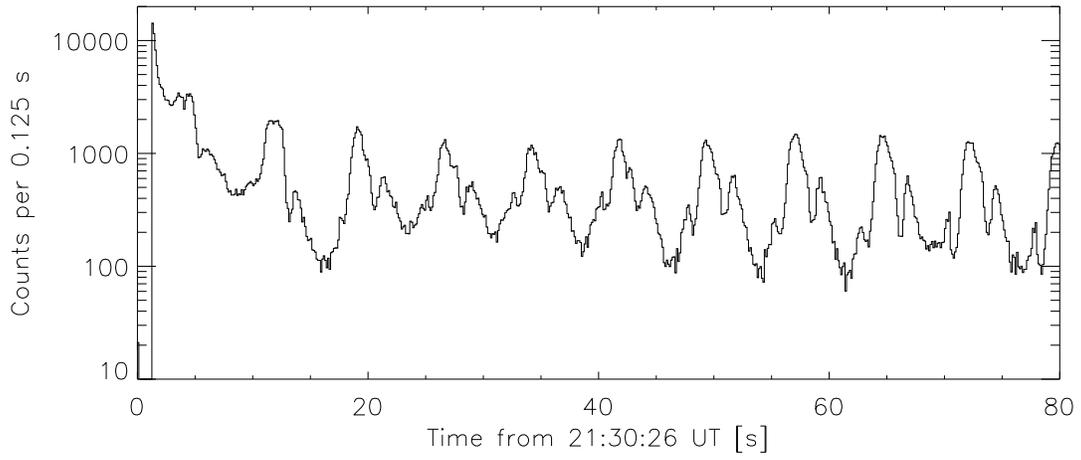}
  \caption{The RHESSI 20-100\,keV time history in 0.125\,s bins as RHESSI comes out of saturation, measuring the main peak decay and the transition to the pulsed tail. \label{fig8}}
  \end{figure}

  \clearpage
  \begin{figure}
  \epsscale{0.80}
  \plotone{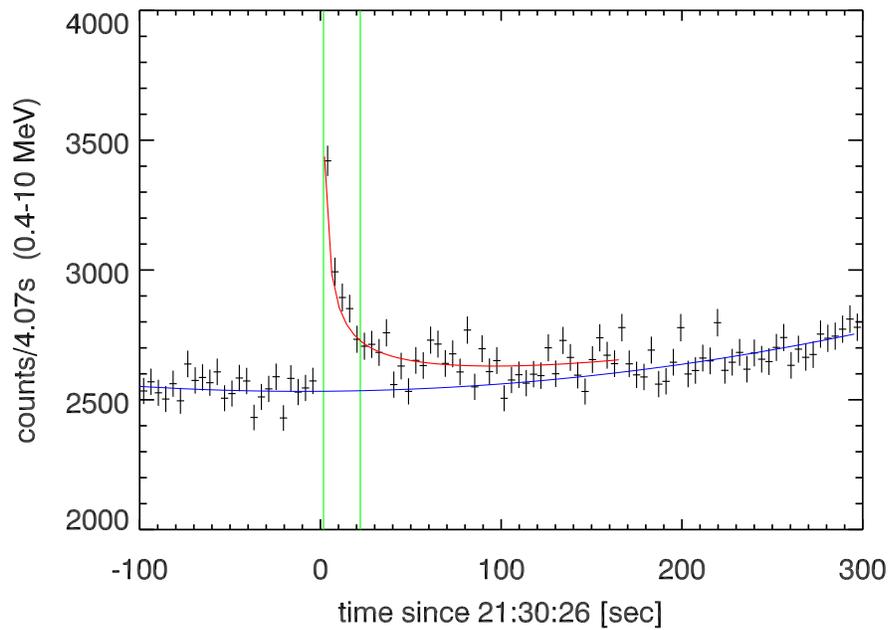}
  \caption{The RHESSI 0.4-10\,MeV lightcurve, 4.07\,s bins (spacecraft rotation period), showing the high energy excess in the decay phase as RHESSI comes out of saturation. The blue curve shows the best fit to the underlying background rate. The red curve shows the best-fit power law decay to the MeV excess rate. The green lines mark the time interval used for spectral analysis (Fig.\,10). \label{fig9}}
  \end{figure}

  \clearpage
  \begin{figure}
  \epsscale{1.00}
  \plotone{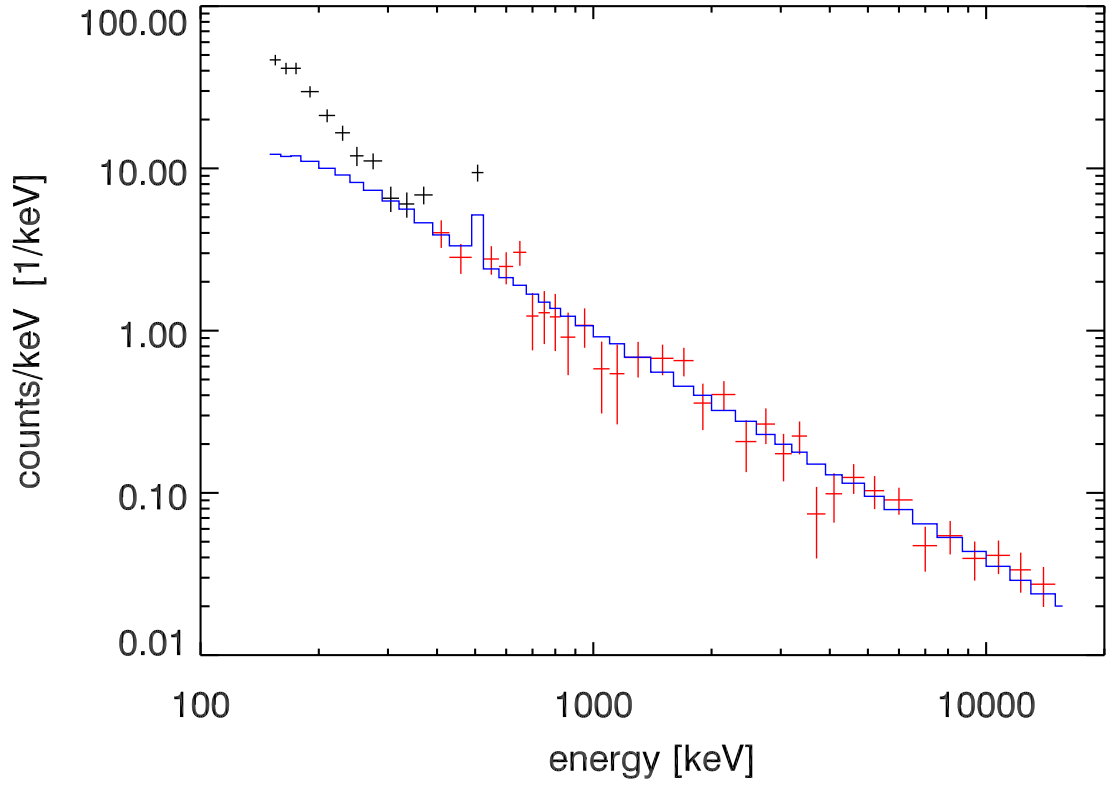}
  \caption{RHESSI MeV spectrum of the high energy excess, integrated over the 20-s interval shown in Fig.\,9. The blue curve shows the best-fit power law model of the 0.4-17\,MeV spectrum (red points). The spectral bin including the 0.511\,MeV background line was excluded from this fit. \label{fig10}}
  \end{figure}

  \clearpage
  \begin{figure}
  \epsscale{0.80}
  \plotone{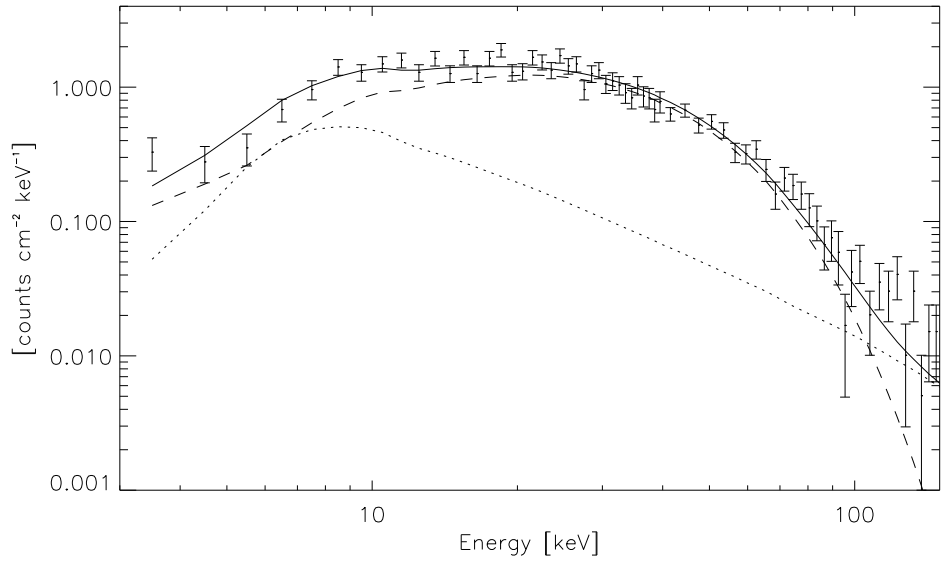}
  \caption{The first RHESSI snapshot spectrum after coming out of saturation, at t$_{26}$ = 2.0\,s. The solid line shows the best-fit absorbed blackbody + powerlaw model convolved through the RHESSI response matrix. The absorbed blackbody (dashed line) and absorbed powerlaw (dotted line) are also shown separately. \label{fig11}}
  \end{figure}

  \clearpage
  \begin{figure}
  \epsscale{0.80}
  \plotone{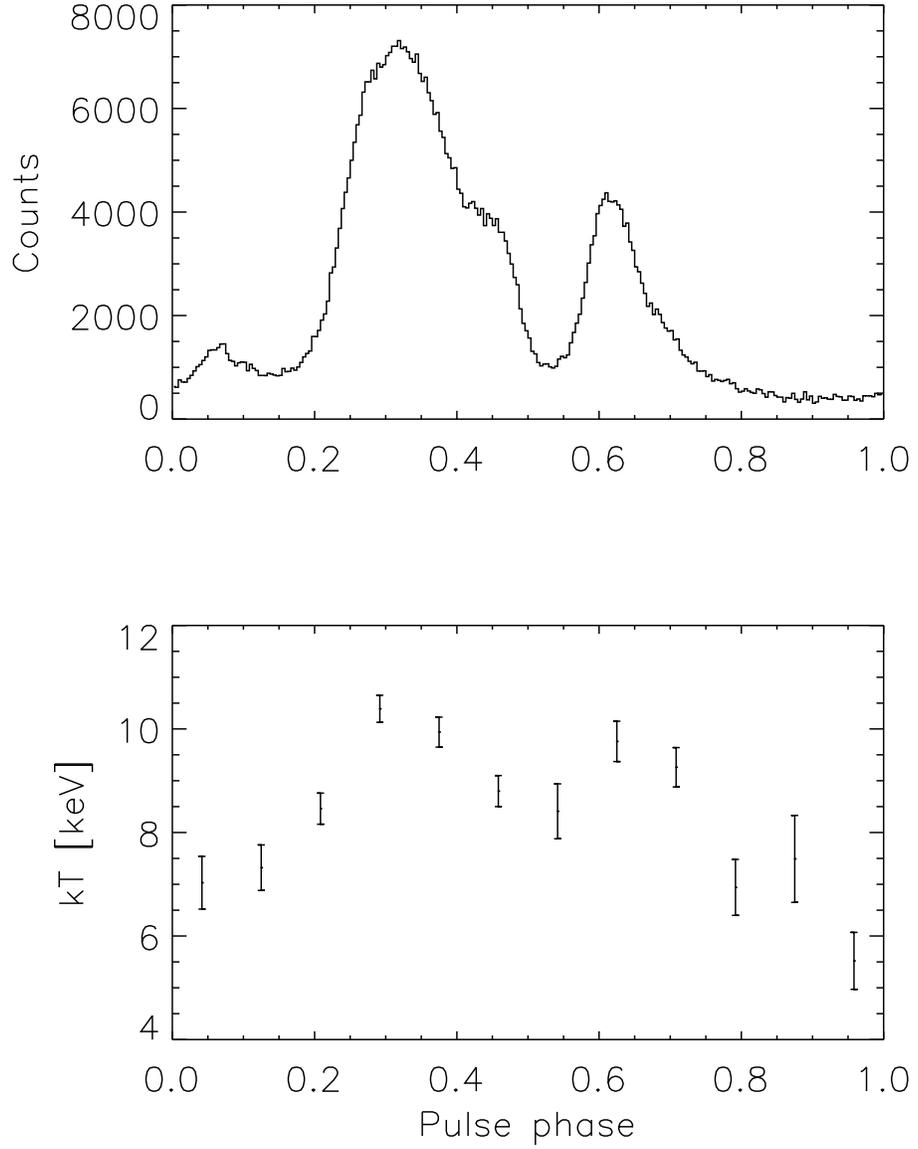}
  \caption{Average pulse shape, 20-100\,keV (top), integrated over the entire 400\,s pulsed tail. The bottom panel shows the phase-resolved blackbody temperature.  \label{fig12}}
  \end{figure}

  \clearpage
  \begin{figure}
  \epsscale{0.60}
  \plotone{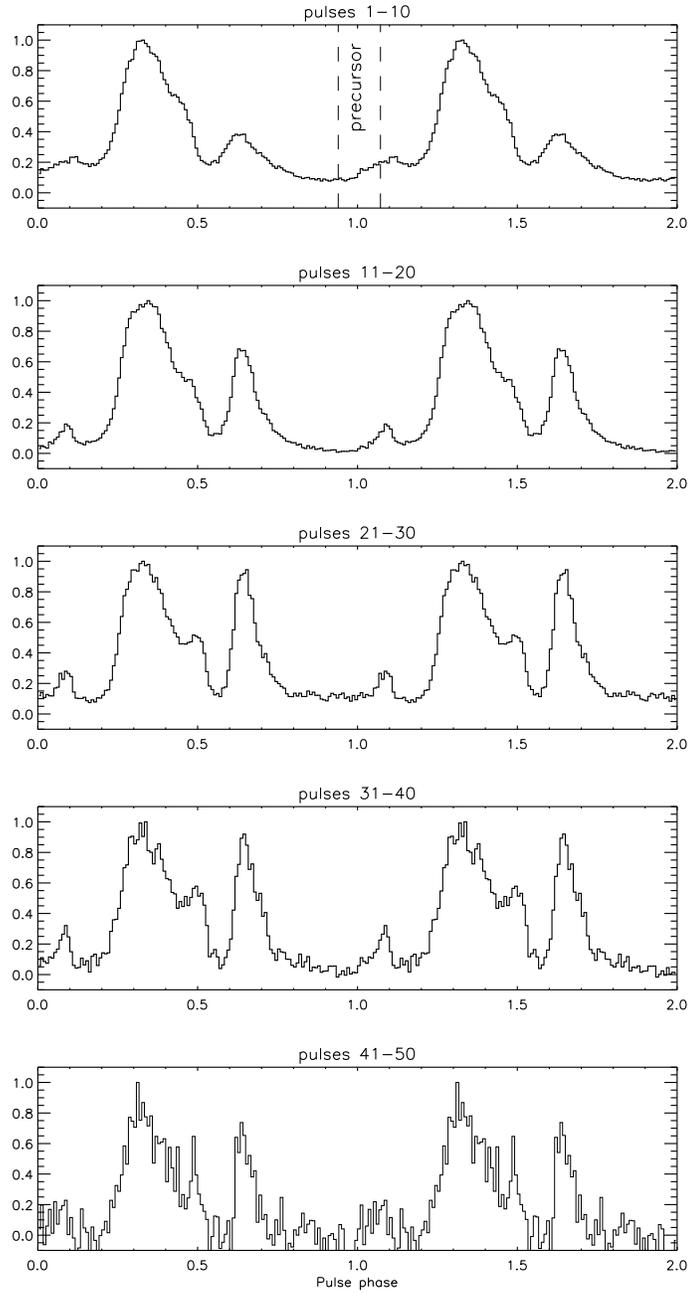}
  \caption{Evolution of the 20-100\,keV pulse profile during the pulsed tail. Each frame shows the pulse profile integrated over 10 successive pulses, from the start of the pulsed tail (top) to the end (bottom). In the top panel, we also show the phase of the precursor relative to the pulsed tail. \label{fig13}}
  \end{figure}

  \clearpage
  \begin{figure}
  \epsscale{0.90}
  \plotone{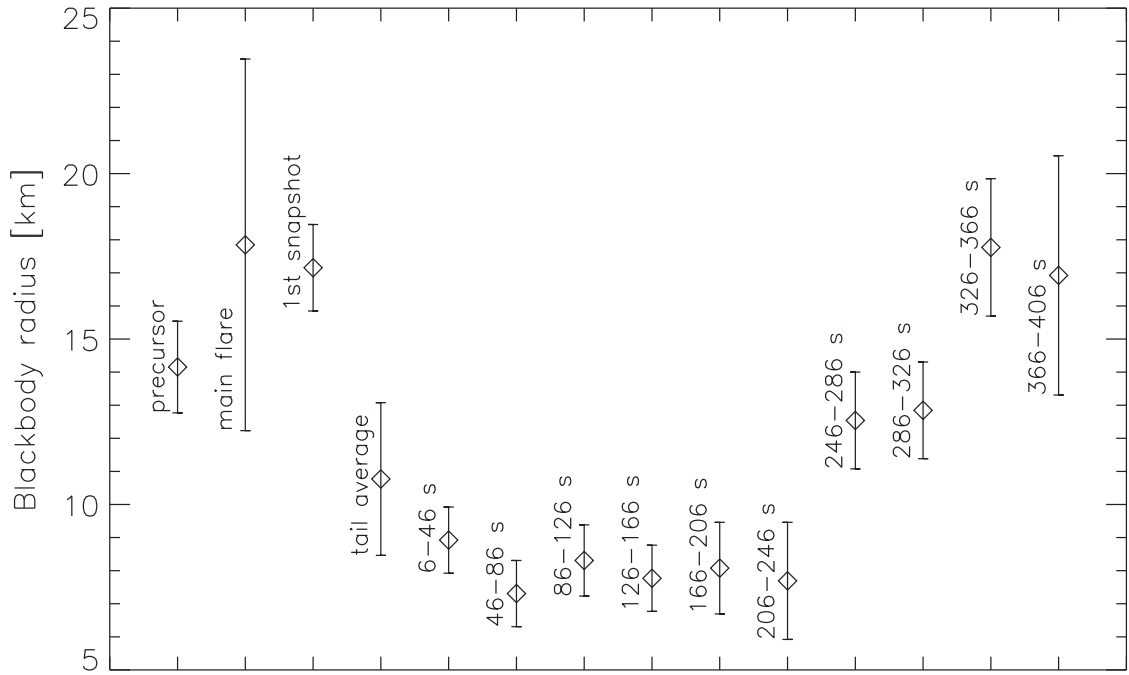}
  \caption{The effective blackbody radius for various stages of the giant flare, assuming a distance of
  10\,kpc and a gravitational redshift of $z = 0.30$. \label{fig14}}
  \end{figure}

  \clearpage

\begin{table}
\begin{center}
\caption{Pulsed tail integrated blackbody + power law spectral fits (3-250\,keV).\label{tbl-1}}

\begin{tabular}{rccccc}
\tableline\tableline
t$_{26}$ & kT$_{bb}$ & BB Fluence\footnotemark[1] & $\Gamma$ & PL Fluence\footnotemark[2] & $\chi_{\nu}^2$\,(dof) \\
 & [keV] & [10$^{-4}$ erg\,cm$^{-2}$] &  & [10$^{-4}$ erg\,cm$^{-2}$] & \\
\tableline
2-6\,s & $11.0 \pm 0.2$ & $3.1 \pm 0.4$ & $1.71 \pm 0.09$ & $0.6 \pm 0.2$ & 1.11\,(73) \\
6-46\,s & $9.7 \pm 0.3$ & $5.0 \pm 0.9$ & $2.34 \pm 0.09$ & $6.3 \pm 2.0$ & 1.23\,(73) \\
46-86\,s & $9.6 \pm 0.4$ & $3.2 \pm 0.7$ & $2.58 \pm 0.14$ & $5.4 \pm 2.2$ & 1.29\,(73) \\
86-126\,s & $9.1 \pm 0.3$ & $3.5 \pm 0.7$ & $2.49 \pm 0.16$ & $3.8 \pm 2.0$ & 0.98\,(73) \\
126-166\,s & $9.4 \pm 0.3$ & $3.4 \pm 0.7$ & $2.64 \pm 0.18$ & $4.4 \pm 2.4$ & 1.01\,(73) \\
166-206\,s & $8.4 \pm 0.4$ & $2.4 \pm 0.7$ & $2.70 \pm 0.15$ & $5.0 \pm 2.5$ & 1.20\,(73) \\
206-246\,s & $7.7 \pm 0.5$ & $1.5 \pm 0.6$ & $2.50 \pm 0.20$ & $2.3 \pm 1.7$ & 1.06\,(73) \\
246-286\,s & $5.9 \pm 0.2$ & $1.3 \pm 0.3$ & $2.1 (fixed)$ & $0.7 \pm 0.1$ & 1.53\,(74) \\
286-326\,s & $5.5 \pm 0.2$ & $1.1 \pm 0.2$ & $2.1 (fixed)$ & $0.6 \pm 0.1$ & 1.16\,(74) \\
326-366\,s & $4.4 \pm 0.2$ & $0.9 \pm 0.2$ & $2.1 (fixed)$ & $0.4 \pm 0.1$ & 0.98\,(74) \\
366-406\,s & $3.5 \pm 0.4$ & $0.3 \pm 0.1$ & $2.1 (fixed)$ & $0.06 \pm 0.07$ & 0.86\,(74) \\
\tableline

\end{tabular}
\end{center}
1. Blackbody fluence integrated over all energies.

2. 3-100\,keV fluence.
\end{table}


\begin{thebibliography}{}

 \bibitem[Cameron et~al.\,(2005)]{cam05} Cameron, P., et al., 2005, \nat, 434, 1112
 
 \bibitem[Corbel et~al.\,(1997)]{cor97} Corbel, S., et al., 1997, \apj, 478, 624

 \bibitem[Corbel \& Eikenberry\,(2004)]{cor04} Corbel, S., and Eikenberry, S., 2004, \aap, 419, 191

\bibitem[Duncan \& Thompson\,(1992)]{dun92} Duncan, R., and Thompson, C. 1992, \apj, 392, L9

 \bibitem[Eikenberry et~al.\,(2004)]{eik04} Eikenberry, S., et al., 2004, \apj, 616, 506

 \bibitem[Feroci et~al.\,(2001)]{fer01} Feroci, M., et al., 2001, \apj, 549, 1021

\bibitem[Frail et~al.\,(1999)]{fra99} Frail, D., Kulkarni, S., and Bloom, J., 1999, \nat, 398, 127

 \bibitem[Gaensler et~al.\,(2005)]{gae05} Gaensler, B. M., et al., 2005, \nat, 434, 1104

 \bibitem[G\"{o}tz et~al.\,(2006)]{got06} G\"{o}tz, D., et al., 2006, \aap, 445, 313

 \bibitem[Hajdas et~al.\,(2005)]{haj05} Hajdas, W., et al., 2005, AIP Conf. Proc., 801, 304

\bibitem[Hurford et~al.\,(2002)]{hur02} Hurford, G. J., et al., 2002, Sol. Phys., 210, 61

\bibitem[Hurley et~al.\,(1999)]{hur99} Hurley, K., et al., 1999, \nat, 397, 41
 
 \bibitem[Hurley et~al.\,(2005)]{hur05} Hurley, K., et al., 2005, \nat, 434, 1098
 
\bibitem[Israel et~al.\,(2005)]{isr05} Israel, G. et al. 2005, \aap, 438, L1

\bibitem[Kosugi et~al.\,(2005)]{kos05} Kosugi, G., Ogasawara, R., and Terada, H. 2005, \apj \, 623, L125

\bibitem[Kouveliotou et~al.\,(1998)]{kou98} Kouveliotou, C., et al. 1998, \nat, 393, 235

\bibitem[Laros et~al.\,(1986)]{lar86} Laros, J., et al. 1986, \nat, 322, 152

 \bibitem[Lin et~al.\,(1995)]{lin95} Lin, R. P., et al., 1995, Space Sci. Rev., 71, 125

 \bibitem[Lin et~al.\,(2002)]{lin02} Lin, R. P., et al., 2002, Sol. Phys., 210, 3

\bibitem[Lorimer \& Xilouris\,(2000)]{lor00} Lorimer, D., and Xilouris, K. 2000, \apj, 545, 385


 \bibitem[Mazets et~al.\,(2005)]{maz05} Mazets, E. P., et al., 2005, astro-ph/0502541

 \bibitem[McClure-Griffiths \& Gaensler\,(2005)]{mcc05} McClure-Griffiths, N., and Gaensler, B., 2005, \apj, 630, L61

 \bibitem[Mereghetti et~al.\,(2005a)]{mer05a} Mereghetti, S., et al., 2005, \apj, 624, L105

 \bibitem[Mereghetti et~al.\,(2005b)]{mer05b} Mereghetti, S., et al., 2005, \apj, 628, 938

\bibitem[Mereghetti et~al.\,(2005c)]{mer05c} Mereghetti, S., Gotz, D., Mirabel, I. 2005, \aap, 433, L9

\bibitem[Molkov et~al.\,(2005)]{mol05} Molkov, S., , et al., 2005, \aap, 433, L13

 \bibitem[Murakami et~al.\,(1994)]{mur94} Murakami, T., et al., 1994, \nat, 368, 127

\bibitem[Paczy\'{n}ski\,(1992)]{pac92} Paczy\'{n}ski, B. 1992, Acta Astronomica 42, 145

 \bibitem[Palmer et~al.\,(2005)]{pal05} Palmer, D. M., et al., 2005, \nat, 434, 1107

 \bibitem[Rea et~al.\,(2005)]{rea05} Rea, N., et al., 2005, \apj, 627, L133

 \bibitem[Schwartz et~al.\,(2005)]{sch05} Rea, N., et al., 2005, \apj, 627, L129

 \bibitem[Smith et~al.\,(2002)]{smi02} Smith, D. M., et al., 2002, Sol. Phys., 210, 33

 \bibitem[Terasawa et~al.\,(2005)]{ter05} Terasawa, T., et al., 2005, \nat, 434, 1110

 \bibitem[Thompson \& Duncan\,(1995)]{tho95} Thompson, C., Duncan, R., 1995, \mnras, 275, 255

 \bibitem[Thompson \& Duncan\,(1996)]{tho96} Thompson, C., and Duncan, R. 1996, \apj, 473, 322

\bibitem[Thompson \& Duncan\,(2001)]{tho01} Thompson, C., Duncan, R., 2001, \apj, 561, 980

\bibitem[Thompson et~al.\,(2002)]{tho02} Thompson, C., Lyutikov, M., and Kulkarnit, S. R., 2002, \apj, 574, 332

\bibitem[Tiengo et~al.\,(2005)]{tie05} Tiengo, A. et al. 2005, \aap, 440, L63

 \bibitem[Woods et~al.\,(2006)]{woo06} Woods, P. M., et al., 2006, \apj, accepted, astro-ph/0602402

  \bibitem[Zehnder et~al.\,(2003)]{zeh03} Zehnder, A., et al., 2003, Proc. SPIE, 4853, 41

  \bibitem[Zoglauer et~al.\,(2006)]{zog06} Zoglauer, A., et al., 2006, New Astron. Rev. 50, 629-632

  \end{thebibliography}
  \end{document}